\documentclass[twocolumn,prd,superscriptaddress,preprintnumbers,floatfix]{revtex4}[10pt]
\pdfoutput=1
\usepackage{amsmath,amssymb,graphicx, multirow} %use drftcite in draftmode
\usepackage{epsf,verbatim}
\usepackage{hyperref}
\usepackage{comment}
\usepackage{cancel}
\usepackage{enumerate} 

\usepackage{color}

\usepackage{array}
\usepackage{rotating}
\usepackage{multirow}

\usepackage{slashbox}

\newcommand{\squishlist}{
 \begin{list}{$\bullet$}
  { \setlength{\itemsep}{0pt}
     \setlength{\parsep}{3pt}
     \setlength{\topsep}{3pt}
     \setlength{\partopsep}{0pt}
     \setlength{\leftmargin}{1.5em}
     \setlength{\labelwidth}{1em}
     \setlength{\labelsep}{0.5em} } }

\newcommand{\squishlisttwo}{
 \begin{list}{$\bullet$}
  { \setlength{\itemsep}{0pt}
     \setlength{\parsep}{0pt}
    \setlength{\topsep}{0pt}
    \setlength{\partopsep}{0pt}
    \setlength{\leftmargin}{2em}
    \setlength{\labelwidth}{1.5em}
    \setlength{\labelsep}{0.5em} } }

\newcommand{\squishend}{
  \end{list}  }

\newcommand{\ifb}{\mathrm{fb}^{-1}}

\newcommand{\gev}{~\mathrm{GeV}}
\newcommand{\tev}{~\mathrm{TeV}}

\newcommand{\fref}[1]{Fig.~\ref{f.#1}}

\newcommand{\sref}[1]{Section \ref{s.#1}}
\newcommand{\ssref}[1]{Section \ref{ss.#1}}

\newcommand{\cref}[1]{Chapter \ref{c.#1}}
\newcommand{\tref}[1]{Table \ref{t.#1}}

\newcommand{\bi}{\begin{itemize} \itemsep=0.75mm}
\newcommand{\ei}{\end{itemize}}
\newcommand{\benum}{\begin{enumerate} \itemsep=0.75mm}
\newcommand{\eenum}{\end{enumerate}}
\newcommand{\ben}{\begin{equation}}
\newcommand{\een}{\end{equation}}
\newcommand{\be}{\begin{equation*}}
\newcommand{\ee}{\end{equation*}}

\begin{document}

\preprint{preprint}

\title{Measuring the $t \bar th$ coupling from SSDL+$2b$ measurements}

\author{David Curtin}
\affiliation{C. N. Yang Institute for Theoretical Physics, Stony Brook University, 
 Stony Brook, NY 11794, U.S.A.}

 \author{Jamison Galloway}
\affiliation{Dipartimento di Fisica, Universita di Roma ``La Sapienza'' \\
{\rm and} INFN Sezione di Roma, I-00185 Rome, Italy}
 
 \author{Jay G. Wacker}
 \affiliation{SLAC, Stanford University, Menlo Park, CA 94025 USA}

\begin{abstract}
We evaluate the potential of a dedicated search for $t \bar t h$ production in the SSDL$+2b$ channel. 
Such a measurement provides direct access to the top Yukawa coupling, since the sensitivity is not convolved with the loop-level $h\gamma\gamma$ 
%or poorly known $h\bar b b$ coupling, 
as is the case for present $t \bar t h$ searches.  Furthermore, susceptibility to uncertainties in the Higgs width can be reduced by considering a ratio of SSDL$+2b$ rates with those of the performed $Wh\rightarrow WWW^*$ measurement.
The SSDL channel can therefore rely  primarily on the already well-measured $h \rightarrow WW^*$ decay.  This feasibility study required the development of a new calculation method for ``fake'' leptons, which constitute the dominant background to our search. Combining measurements from LHC7, LHC8 and in the future LHC14 for the Higgs coupling fit would help resolve any remaining ambiguity between the top Yukawa coupling and a BSM contribution to the $hgg$ coupling. 
\end{abstract}

\maketitle

 \setcounter{equation}{0} \setcounter{footnote}{0}

 %%%%%%%%%%%%%%%%%%%%%%%%%%%%%%%%%%%%%%%%%%%%%%%%
 %%%%%%%%%%%%%%%%%%%%%%%%%%%%%%%%%%%%%%%%%%%%%%%%
\section{Introduction}
\label{s.intro} 
 %%%%%%%%%%%%%%%%%%%%%%%%%%%%%%%%%%%%%%%%%%%%%%%%
 %%%%%%%%%%%%%%%%%%%%%%%%%%%%%%%%%%%%%%%%%%%%%%%%

The recently discovered scalar boson  \cite{ATLASdisc, CMSdisc} appears to have properties that qualitatively match those predicted by the Standard Model Higgs  \cite{ATLASprop, CMSprop}. In this post-discovery era verifying the precision predictions of the Higgs is of paramount importance. This requires as many measurements of different Higgs production and decay combinations as possible. 

One of the most important Higgs couplings is also the largest, namely the Yukawa coupling to top quarks. Unfortunately, presently performed searches provide no direct measurement of this critical parameter. This article shows that it is possible to measure the top Yukawa coupling using the $WW^*$ decay mode of the Higgs boson in top-Higgs associated production: $pp\rightarrow t\bar{t} H \rightarrow (bW^+)(\bar{b}W^-) (W^\pm W^{\mp *})$. This measurement has important advantages compared to existing searches of $t \bar t h$ production, which rely either on a loop coupling in the Higgs decay that is sensitive to quantum effects ($h \rightarrow \gamma \gamma$ \cite{cmstthgaga}) or an additional poorly constrained Yukawa coupling and a relatively low signal to background ratio ($h \rightarrow b \bar b$ \cite{cmstthbb, atlastthbb}). The tree-level $hWW$ coupling, on the other hand, is well constrained by other measurements. Therefore a search of comparable sensitivity will provide a better measurement of the top Yukawa coupling, after taking the signal strength ratio with a performed $Wh \rightarrow WWW^*$ measurement \cite{whtowww} to eliminate susceptibility to total higgs width uncertainties (mostly from the poorly known bottom Yukawa coupling).  

This search was first proposed over a decade ago by~\cite{oldtthww}. Since then, this important channel has not been re-examined in light of current LHC measurements and capabilities. Furthermore, \cite{oldtthww} did not include the fake lepton contribution, which is actually the dominant source of background as we discuss below. 

Existing searches for physics beyond the Standard Model are sensitive to the  $t\bar{t}W^+W^-$ final state, but because they have not yet been optimized to the Higgs channel, they do not provide competitive limits.  However, modest changes to the search regions will provide competitive limits on the top quark Yukawa coupling.  This improved limit will have an effect in resolving the tension between the ATLAS and CMS inferred measurements of the top Yukawa coupling.  Using just the LHC7 and LHC8 data, fits of the top Yukawa will get dramatically closer to the Standard Model for ATLAS by the inclusion of this one measurement.  In order to demonstrate the potential gain and estimate its size, the existing limit from the CMS same sign dilepton (SSDL) is first calculated.  This requires simply computing the signal efficiency and using the stated background measurements in signal regions. and therefore it is possible to do with relatively high confidence.  The rest of the article will describe how we extrapolated measurements into signal regions that have not been explicitly presented.

\subsubsection*{Current Experimental Limit}

The SSDL $+ 2 b$ searches performed by CMS with 5$\ifb$ at LHC7 \cite{cmsSSDL2b} and 10$\ifb$ at LHC8 \cite{cmsSSDL2bLHC8} are a good starting point to look for evidence of $t \bar t h$ production in the SSDL channel. (The SSDL$+b$ searches performed by ATLAS \cite{atlasSSDL} use either 1 or 3+ $b$-tags and are not as well suited to this purpose.) 
It is a simple matter to calculate the $t \bar t h$ signal and obtain a signal strength exclusion from the number of expected signal events. (See \sref{collider} for technical details.) The CMS search defines different SSDL signal regions in the ($\cancel{E}_T$,$H_T$)-plane, but since the small $t \bar t h$ signal has such similar kinematics to the background these variables are not well-suited for signal isolation. The best exclusion limit is derived from their results in ``Signal Region 0" ($H_T > 80 \gev$, no $\cancel{E}_T$ requirement). Combining the two searches gives a signal strength limit of
\begin{eqnarray}
\mu_{t\bar th}(b\bar{b}\ell^\pm\ell^\pm) &<& 8.7 \; (9.2)
\end{eqnarray}
for 10\% (20\%) systematic uncertainty on the signal efficiency, where 
\begin{eqnarray}
\mu_{t\bar th}(b\bar{b}\ell^\pm\ell^\pm) &\equiv& \frac{ \sigma(t \bar t h) \times \mathrm{Br}(t \bar t h \rightarrow b\bar{b}\ell^\pm\ell^\pm) \  \ \ }{[\sigma(t \bar t h) \times \mathrm{Br}(t \bar t h \rightarrow b\bar{b}\ell^\pm\ell^\pm)]_\mathrm{SM}}.
\end{eqnarray}
Approximately 25\% of SSDL events passing cuts actually originate from $h\rightarrow \tau \tau$ decays. Since the $h\tau\tau$ coupling is already determined to be SM-like with 40\% precision by the combined LHC7 and LHC8 data \cite{htautau}, this uncertainty enters the top coupling determination at the 10\%-level and we will therefore assume it to be of SM strength in what follows.  In this case we write
\begin{eqnarray}
 \mu_{t\bar th}(b\bar{b}\ell^\pm\ell^\pm) & \approx & c_t^2 c_V^2 \times \gamma 
 %\approx \  c_t^2 \times \frac{ \Gamma_{h}^{\rm (SM)}}{\Gamma_h \ \ \ } .
\end{eqnarray}
with $c_t$ and $c_V$ we denote the couplings $g_{htt}$ and $g_{hVV}$ respectively, both as normalized by their SM values and
\begin{eqnarray}
\gamma\equiv \frac{ \Gamma_{h}^{\rm (SM)}}{\Gamma_h \ \ \ } .
\end{eqnarray}
%%
%The $\gamma$ dependence can be eliminated by forming a ratio of another signal strength.    The signal strength measurement of $Wh\rightarrow WWW^*\rightarrow \ell\ell'\ell''$ \cite{whtowww},  more generally $Vh\rightarrow VV'V'^*$,  has the property 
%%
Dependence on $\gamma$ (and hence the poorly known bottom Yukawa) can be eliminated by making use of a $Vh\rightarrow V V' V'^{*}$ measurement, where the signal strength scales as 
\begin{eqnarray}
\mu_{Vh\rightarrow VV'V'^*} \approx c_V^4\times \gamma.
\end{eqnarray}
Therefore, the top coupling can be measured without introducing any new couplings through the ratio
\begin{eqnarray}
\frac{ \mu_{t\bar th}(b\bar{b}\ell^\pm\ell^\pm)}{ \mu_{Wh}(\ell\ell'\ell'')} = \frac{ c_t^2}{ c_V^2} .
\end{eqnarray}

The $Vh \rightarrow$ trilepton measure has similar or better sensitivity than the $t\bar{t}h\rightarrow b\bar{b}\ell^\pm\ell^\pm$ channel \cite{whtowww} and will reach Standard Model sensitivity by the time that the time that $\mu_{t\bar{t}h}$ reaches the Standard Model sensitivity.  
%Since $c_V\le 1$ in theories without exotic Higgs representations, this will cap the flat direction $c_V\rightarrow \infty$ with $c_t/c_V$ fixed.   
Additionally, $\mu_{qqh\rightarrow qqVV^*}$ will provide a similar measurement proportional to $c_V^4\times \gamma $. 
Since the coupling $g_{hVV}$, and hence $c_V$,  is already well-determined to be SM-like (see e.g. \cite{Falkowski}), there is direct access to $c_t$.  This illustrates the advantage of the $b\bar{b}\ell^\pm\ell^\pm$ channel in measuring $c_t$.  
We will assume for this study that $c_V^4\times\gamma$  is measured sufficiently well to not introduce significant error which is equivalent to
assuming for the time being that $\gamma= c_V = 1$.

The goal of this article is to demonstrate that sensitivity to the top coupling can be significantly improved in a dedicated SSDL$+2b$ search with cuts optimized to the $t \bar t h$ signal.  This requires being able to extrapolate Standard Model backgrounds into regions that have not been directly searched for at the LHC.  For irreducible backgrounds such as $t\bar{t} W^\pm$, this is a relatively straightforward procedure of calculating backgrounds using Monte Carlo generators.   However, the irreducible backgrounds are not the dominant backgrounds in these searches.  Instead,  many of the SSDL background events arise from ``fake'' leptons, which are leptons that are not generated at the matrix element level or in the subsequent decays of $W^\pm$ and $Z^0$ vector bosons or top quarks.   As an example, about half of the total background at $\sqrt{s} = 7 \tev$ is believed to have come from fake leptons, and this contribution becomes increasingly dominant at higher center-of-mass energies. 

Traditionally fake leptons are considered beyond the reach of estimation.  In this article, a new method of calculating the rate of fake leptons is devised.  This method makes use of well-understood Monte Carlo for the processes which `source' the fake signal. These sources are convoluted with a set of transfer functions and efficiency functions that are grounded in the physical processes which give rise to fake leptons.  The parameters in these transfer and efficiency functions are then fit to existing experimental results.  This procedure makes the problem of calculating lepton fake rates computationally tractable. The Fast FakeSim will be explored and developed in-depth in \cite{fakesim}, and  we provide a summary of the needed details in \sref{fakesim}.

With the Fast FakeSim in hand we undertake a $t \bar t h$ SSDL collider study in \sref{collider} to optimize the exclusion attainable with current data, and extrapolate to the kinds of analyses that could be undertaken at the 14 TeV LHC to achieve SM-like sensitivities with 30 $\ifb$. We find that the optimal cuts  are significantly different from those employed by \cite{cmsSSDL2b, cmsSSDL2bLHC8}. Our results are largely conservative -- experimental improvements beyond the scope of our study, like tightened lepton-ID or multivariate techniques, should be able to produce significant improvements on the sensitivities we demonstrate. Our study is intended to serve as a proof-of-feasibility for the SSDL $t \bar t h$ measurement, as well as motivation for experimental work on reducing systematics related to fake lepton background.

In \sref{Higgs} we examine the impact of the present and future possible $t \bar t h$ measurements in the SSDL channel on determinations of the Higgs couplings to both tops and gluons. We find that our proposed measurements can help resolve a possible degeneracy between these two couplings. The top Yukawa coupling itself can be measured with $\mathcal{O}(10\%)$ precision. We summarize our findings in \sref{conclusions}.

% or make this section 1.1
% section 2: fakesim --> fold in atlas etc
% section 3: collider analysis
%% flavor separation
% section 4: Higgs couplings
% conclusions

 %%%%%%%%%%%%%%%%%%%%%%%%%%%%%%%%%%%%%%%%%%%%%%%%
 %%%%%%%%%%%%%%%%%%%%%%%%%%%%%%%%%%%%%%%%%%%%%%%%
\section{Simulation of Fake Leptons}
\label{s.fakesim}
 %%%%%%%%%%%%%%%%%%%%%%%%%%%%%%%%%%%%%%%%%%%%%%%%
 %%%%%%%%%%%%%%%%%%%%%%%%%%%%%%%%%%%%%%%%%%%%%%%%
Fake leptons constitute a dominant source of background for any search in the SSDL final state, and accounting for their contribution was the main technical obstacle to performing our $ t \bar t h$ discovery study. 

`Fake' in this context can mean two things: \emph{irreducible fakes} are real electrons or muons produced mostly from heavy flavor decays inside of jets; \emph{reducible fakes} are thin hadronic jets that fake an electron signature inside the detector. (Reducible muon fakes are negligible.) Irreducible lepton fakes can in principle be calculated with Monte Carlo generators, but the rates are of order $10^{-4}$ per QCD jet, which makes it impractically difficult to achieve a statistically meaningful sample. 
Simulating reducible lepton fakes requires full detector simulation. This is is extremely challenging and  beyond the scope of any  theoretical study.  For this reason, data-driven methods are typically used to crudely estimate the fake lepton contributions \cite{cmsSSDL2j, cmsSSDL2b}.

For our $t\bar t h$ study we devised a new method to calculate lepton fake rates. This method is computationally efficient, requiring the generation of only $\sim$ 10,000 events per source process. It is grounded in physical principles but tuned to and verified by data. 

Fake electrons arise from jets and there is relationship between the originating jet and the fake electron.
With a large fake electron sample and truth-level event information one could  construct two functions: a \emph{mistag efficiency}  $\epsilon_{j \rightarrow e} = \epsilon_{j \rightarrow e}(j_\mathrm{orig})$, which represents the probability that a particular jet $j_\mathrm{orig}$ somehow creates a fake electron; and the \emph{transfer function} $\mathcal{T}_{j \rightarrow e}  = \mathcal{T}_{j \rightarrow e} (j_\mathrm{orig}; e_\mathrm{fake})$, which represents a probability distribution of possibilities for the properties of the fake electron (normalized to unity), which will also depend on the properties of the original jet $j_\mathrm{orig}$.  If the efficiency and transfer functions could be accurately determined, then the fake electron contribution to a particular process could be easily calculated by convoluting those functions with a relatively small event sample for the processes which \emph{source} the fake electron background for our particular study.  (The same discussion applies to fake muons or other fake detector objects.) 

In the present study, most of the fake lepton background for the SSDL + $2b$ final state comes from $t \bar t$ and $Wbbj$ with one $W$ giving a real electron and one of the light jets faking another, as well as dileptonic $t \bar t + j$ with the jet faking a third lepton.

The challenge of determining these efficiency and transfer functions can be  simplified with a few physical approximations. 
If our study only uses detector objects within the tracker of the ATLAS or CMS experiments (i.e. $|\eta| < 2.5$), then the efficiency curve can be parameterized as a one-dimensional function of original jet $p_T$ for a given set of lepton isolation conditions. Furthermore, for reasonably hard jets with $p_T \gg m_b$, any heavy flavor decay will be sufficiently boosted so that the fake electron will be roughly collinear with the original jet. Reducible fakes are also aligned. In order for the `lepton' to be hard enough to be detected and pass isolation conditions, the remainder of the original jet (a few hadronic tracks and/or  a neutrino from heavy flavor decay, if anything) has to be soft. In that case, the transfer function can be simplified to a one-dimensional PDF for the parameter $\alpha \in (0, 1)$, which represents the fraction of the original jet $\vec{p}$ that is `lost' when converting to a fake lepton. The remaining balance of the jet is assumed to turn into MET, and we assume that the momentum loss scales with jet $\vec{p}$, i.e. $\mathcal{T}(\alpha)$ does not depend on the properties of the original jet. Finally, we have to chose particular parameterizations for the functions $\epsilon_{j \rightarrow \ell}(p_{T_j})$ and $\mathcal{T}_{j \rightarrow \ell}(\alpha)$. For the efficiency curve, public experimental data on mistag rates (e.g.\cite{effcurvelinear}) suggests a simple linear function in $p_T$ to be a reasonable choice in our regime of interest; for the transfer function we will assume a simple gaussian with mean $\mu$ and spread $\sigma$, normalized to unit-area in the interval $\alpha \in (0,1)$:
\begin{eqnarray}
\nonumber \epsilon_{j\rightarrow \ell}(p_{T_j}) & = & \epsilon_{200} \left[ 1 - (1-r_{10}) \frac{200 - (p_{T_j} / \gev) }{200 - 10}\right]\\
\nonumber \mathcal{T}_{j\rightarrow \ell}(\alpha) &=& \frac{ 
\frac{1}{\sqrt{2 \pi} \sigma} e^{- \frac{(\alpha - \mu)^2}{2 \sigma^2}}
}{
\frac{1}{2} \left[ \mathrm{Erf}\left(\frac{1-\mu}{\sqrt{2}\sigma}\right) + \mathrm{Erf}\left(\frac{\mu}{\sqrt{2}\sigma}\right)\right]}\\
\nonumber \mathrm{where} \ \ \ \ \alpha &\equiv& 1 - \frac{\vec{p}_{e_\mathrm{fake}}}{\vec{p}_{j_\mathrm{orig}}},\\
\nonumber \epsilon_{200} & \equiv & \epsilon_{j\rightarrow \ell}(200 \gev), \\
 r_{10}& \equiv &\frac{ \epsilon_{j\rightarrow \ell}(10 \gev)}{ \epsilon_{j\rightarrow \ell}(200 \gev)}.
\end{eqnarray}
While these approximations regarding fake leptons are physically motivated  they have to be verified with experimental data. Assuming for simplicity \footnote{This is justified if our final observables are lepton flavor blind.} that $\epsilon_{j\rightarrow e} = \epsilon_{j \rightarrow \mu}$, $\mathcal{T}_{j \rightarrow e} = \mathcal{T}_{j \rightarrow \mu}$ we then have to determine four parameters, which can be fixed by reproducing the experimental estimates for the fake lepton background in the signal regions of the CMS LHC7 $5\ifb$ SSDL+$2b$ search \cite{cmsSSDL2b}.

Our fake lepton simulation method (FakeSim) reproduces the result from \cite{cmsSSDL2b} remarkably well. Scanning over  $(\mu, \sigma, r_{10}, \epsilon_{200})$ we find that the transfer function parameters are constrained to be $\sigma \approx 0.1, \mu \lesssim 0.1$, which lends support to our procedure since it independently reproduces the public CDF results on fake lepton $E_T$ as a function of originating jet $E_T$ \cite{cdftransferfnmeasurement}.  As for the efficiency curve, for a given choice of other parameters $\epsilon_{200}$ is fixed by the overall fake lepton count in \cite{cmsSSDL2b}, but  the gradient $r_{10}$ is less well constrained by the available information. This is not entirely unexpected -- the fake leptons from our source processes are sensitive to some $p_T$-region of the efficiency curve which is constrained by the fit, with only the high-momentum tails sensitive to the curve's gradient. Nevertheless, what public data on electron mistag efficiency we had available (e.g. \cite{effcurvelinear}) was reproduced very well by one of the best-fit parameter points ($r_{10} = 0$ in table below), up to an overall $\mathcal{O}(1)$ rescaling which could be due to our equal treatment of electrons and muons, or simply due to actual differences in the mistag rate for our source sample.

Due to this ambiguity in the fit we use three possibilities for the FakeSim parameters in our $t \bar t h$ study, representing the spread of possibilities allowed by reproducing the fake estimates from  \cite{cmsSSDL2b}. They are compared to the experimental prediction in \fref{fakesim}, and their explicit values are given in \tref{fitparams}

Allowing for these different possibilities, our predictions for the numbers of fake lepton events in the signal regions of \ssref{lhc78} vary by 10\%, which can be understood as the systematic error of our FakeSim output from requiring agreement with the predictions of \cite{cmsSSDL2b}. Since those predictions  themselves quoted a 50\% systematic error we can safely neglect this additional systematic on the overall background expectation. Furthermore, the relatively small dependence of the outcome on the precise FakeSim parameters also implies that the main physics of fake leptons is captured by the Monte Carlo calculation of the source processes like $t \bar t$ and $Wbbj$ production, and by the correct transfer function which arises mostly from heavy flavor decay kinematics and the requirement of passing lepton isolation conditions.

CMS released a $10\ifb$ LHC8 update on their SSDL+$2b$ search \cite{cmsSSDL2bLHC8} with identical cuts to \cite{cmsSSDL2b} except for slightly modified lepton reconstruction.  The FakeSim with the parameters of \tref{fitparams} reproduced the fake lepton expectations in the various signal regions of the LHC8 search very well if the overall mistag rate was uniformly scaled up by 30\%. This increase may be due to pile-up degrading lepton identification, or due to the aforementioned changes in lepton reconstruction. The modest upscaling does not detract from the excellent shape agreement of the FakeSim, which serves as independent verification of our fake simulation principles and increases confidence in our study at $\sqrt{s} = 8 \tev$, performed with the stated $\epsilon_{200}$ increase.

\begin{figure}
\begin{center}
\begin{tabular}{c}
CMS LHC7, 5$\ifb$\\
\includegraphics[width=8cm]{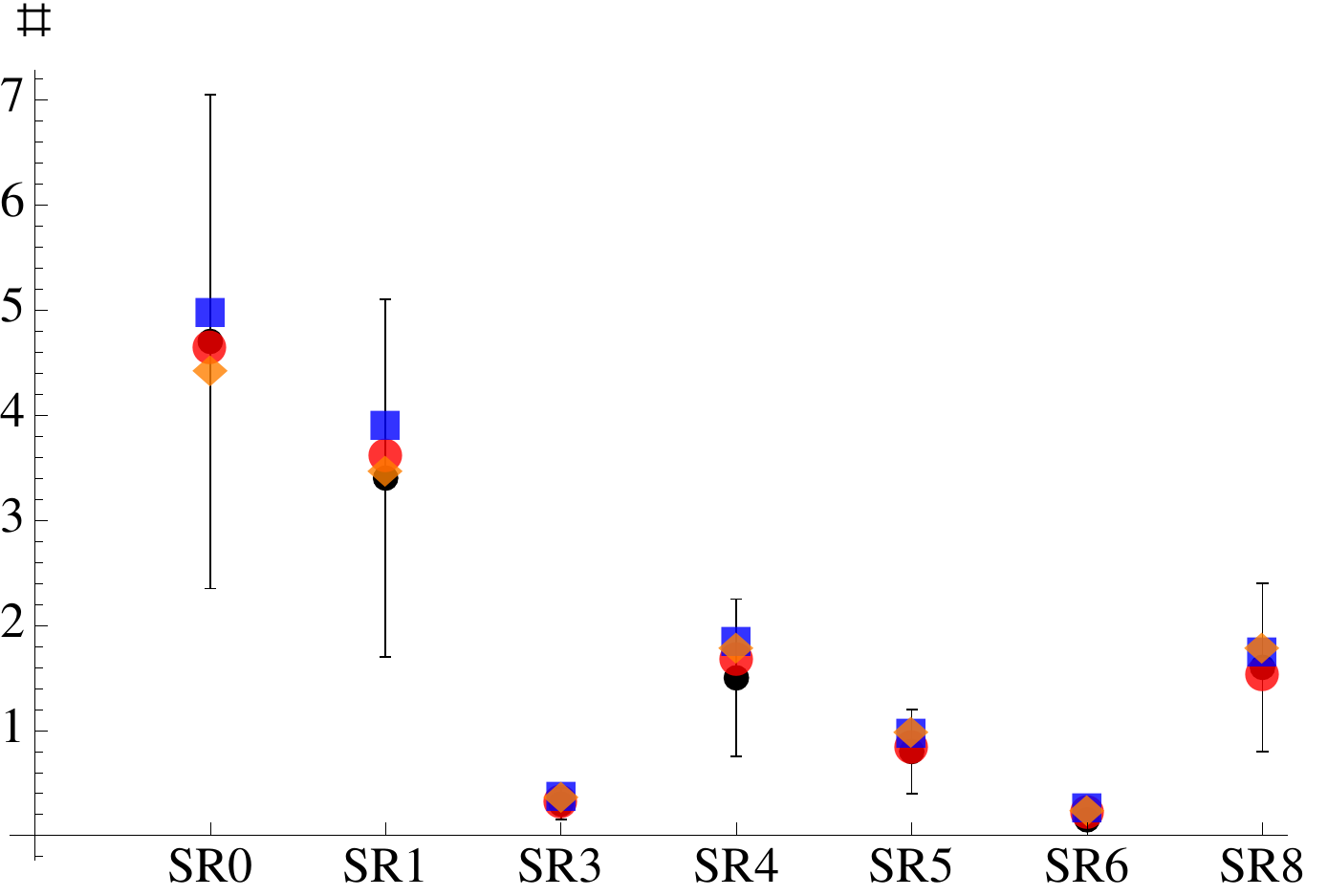}
\vspace{3mm}
\\
CMS LHC8, 10$\ifb$\\
\includegraphics[width=8cm]{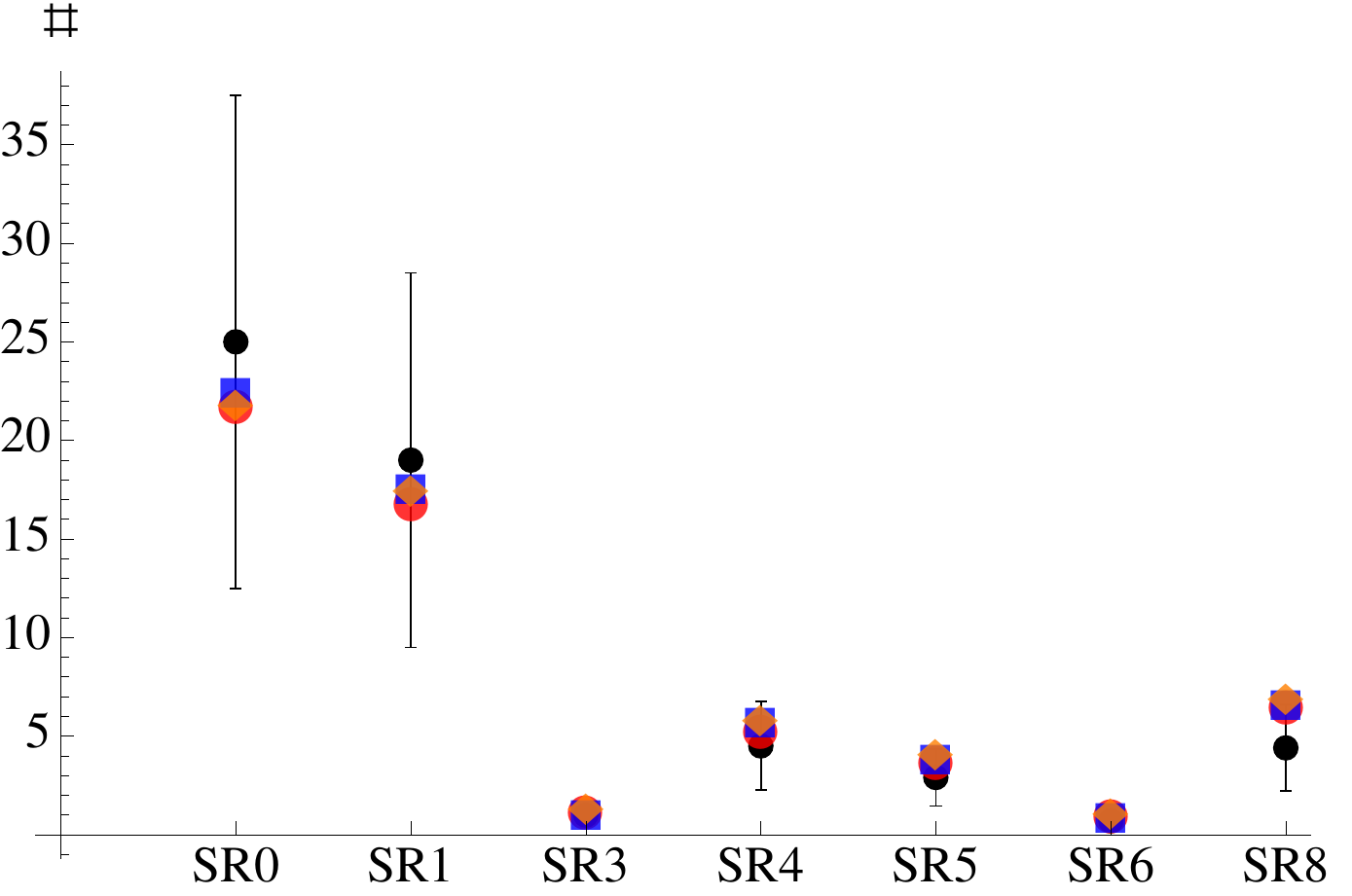}
\end{tabular}
\end{center}
\caption{
Prediction for the number of fake lepton events in the Signal Regions of the CMS SSDL+$2b$ searches \cite{cmsSSDL2b, cmsSSDL2bLHC8} at LHC7 and LHC8. Black: CMS predictions, shown with 50\% systematic error bars. Orange Diamonds, Blue Squares,  Red Circles: predictions of our Fast FakeSim for the fit parameters in \tref{fitparams} with $r_{10} = 0, 0.5, 1$. The FakeSim was tuned to reproduce the LHC7 case with a parameter scan in $(\mu, \sigma, r_{10}, \epsilon_{200})$. The successful reproduction of the LHC8 experimental prediction serves as a  validation of our methods. (An overall increase in the fake lepton rate for LHC8 is compensated for by a 30\% upscaling of $\epsilon_{200}$ compared to LHC7.)
}
\label{f.fakesim}
\end{figure}

\begin{table}
\begin{center}
\begin{equation}
\begin{array}{|c|c|c|c|}
\hline
\mu & \sigma &  r_{10} &  \epsilon_{200} \\
\hline
0  & 0.1  & 0   &2.9 \times 10^{-4} \\
0 & 0.15 & 0.5  &1.2 \times 10^{-4} \\ 
 0 & 0.05 & 1  &0.65 \times 10^{-4}\\ \hline
\end{array}
\end{equation}
\end{center}
\caption{The three parameter choices used in our Fast FakeSim, representative of the spread of values allowed by maximizing agreement with \cite{cmsSSDL2b}.
}
\label{t.fitparams}
\end{table}

For our LHC14 collider study there was no calibration information available, so we examined two different scenarios for the fake lepton contribution, either a fake rate that is 30\% higher than LHC7 (i.e. same as LHC8) or 100\% higher. The larger amount of signal at LHC14 favor tighter cuts to optimize the $t \bar t h$ sensitivity, which makes the limit sensitive to kinematic regions where the FakeSim has not been very well-constrained by current data. This results in $\mathcal{O}(50\%)$ differences in the fake lepton predictions. We refer to the $r_{10} = 0$ ($r_{10} = 1$) fake lepton prediction from \tref{fitparams}  with a 30\% (100\%) upscaling of $\epsilon_{200}$ as the optimistic (pessimistic) scenario, and evaluate $t \bar t h$ exclusions for both to explore the range of possible outcomes at the LHC14. We place slightly higher trust in the optimistic prediction since the $r_{10} = 0$ set of fit parameters reproduces the experimentally determined electron mistag curve \cite{effcurvelinear} so well. 
This procedure does not take into account possible improvements that can be implemented by the experimental collaborations, such as tighter lepton-ID or more sophisticated multivariate discriminators. 

It is also conceivable that our FakeSim, with additional experimental data and verification, might help to reduce fake lepton predictions and systematic uncertainties for the experimental studies themselves. This is because it makes maximal use of well-understood Monte Carlo for the processes which source fake leptons. The remainder of the problem is expressed in terms of tractable transfer and efficiency functions, which are amenable to theoretical and Monte-Carlo exploration (for example, deriving $\epsilon, \mathcal{T}$ from hadron decay kinematics and splitting functions) as well as dedicated experimental tuning studies. 

The current incarnation of the Fast FakeSim serves as a proof-of-concept. The ideas presented in this paper will be further developed in \cite{fakesim}, where issues like fake rates for heavy flavor jets and reproduction of differential fake distributions using ATLAS fake lepton predictions \cite{atlasSSDL}
(in both kinematic variables and jet multiplicity) will be addressed.

 %%%%%%%%%%%%%%%%%%%%%%%%%%%%%%%%%%%%%%%%%%%%%%%%
 %%%%%%%%%%%%%%%%%%%%%%%%%%%%%%%%%%%%%%%%%%%%%%%%
\section{Collider Analysis}
\label{s.collider} 
 %%%%%%%%%%%%%%%%%%%%%%%%%%%%%%%%%%%%%%%%%%%%%%%%
 %%%%%%%%%%%%%%%%%%%%%%%%%%%%%%%%%%%%%%%%%%%%%%%%
 
 We will now show that a search in the SSDL+$2b$ final state is sensitive to a $t \bar t h$ signal, where $h \rightarrow WW^*$ or, about 25\% of the time for SM-like $y_\tau$, $h\rightarrow \tau \tau$. We treat each existing dataset at the LHC in turn before discussing the outlook at the 14 TeV LHC.

\begin{table*}
\begin{center}
\begin{tabular}{|l|l||l|l|}
\hline
\multicolumn{4}{|c|}{\emph{Common Preselection}}\\
\hline
\multicolumn{4}{|l|}{Two $b$-jets with $p_T> 40 \gev$, $|\eta| < 2.4$}\\
\multicolumn{4}{|l|}{Two SS leptons with $p_T > 20 \gev$, $|\eta| < 2.4$} \\
\multicolumn{4}{|l|}{$Z$-veto for events with 3+ leptons}\\
\hline 
\emph{LHC7, LHC8 5$\ifb$} & \emph{LHC8 20$\ifb$} & \multicolumn{2} {c|}{\emph{LHC14 30$\ifb$}}\\
\hline
$m_\mathrm{SSDL} < 175 \gev$  & $m_\mathrm{SSDL} < 175 \gev$  & \multicolumn{2}{l|}{$m_\mathrm{SSDL} < 110 \gev$} \\
OR & OR &  \multicolumn{2}{l|}{$ p_{T_{\ell1}} < 110 \gev$} \\
$\big(\ p_{T_{\ell 1}} < 250 \gev$ AND & $\big(\ p_{T_{\ell 1}} < 250 \gev$ AND  
&\multicolumn{2}{l|}{Require 3$^\mathrm{rd}$ lepton} \\
$\ \ \Delta R_\mathrm{SSDL} < \pi\big)$&$ \ \ $ 3$^\mathrm{rd}$ lepton $\big)$ &
\multicolumn{2}{l|}{ } \\
\hline
\end{tabular}
\end{center}
\caption{
Summary of cuts used in our analyses. The common preselection and reconstruction criteria are identical to those of \cite{cmsSSDL2b, cmsSSDL2bLHC8}
}
\label{t.cuts}
\end{table*}

  %%%%%%%%%%%%%%%%%%%%%%%%%%%%%%%%%%%%%%%%%%%%%%%%
 \subsubsection{LHC with $\sqrt{s} = 7 $ and $ 8 \tev$}
 %%%%%%%%%%%%%%%%%%%%%%%%%%%%%%%%%%%%%%%%%%%%%%%%
\label{ss.lhc78}

Our $t \bar t h$ study was inspired by the SSDL + 2$b$ search performed by CMS with $5\ifb$ of data at LHC7 \cite{cmsSSDL2b}, and uses the same cuts as a preselection: two same-sign dileptons with $p_T > 20 \gev$ and two $b$-tagged jets with $p_T > 40 \gev$, both with $|\eta| < 2.4$. Events with more than two leptons are vetoed if a $Z$-candidate satisfying $|m_{\ell\ell} - m_Z| < 15 \gev$ can be reconstructed. 

At the LHC, $t\bar t W$ and $t \bar t Z$ production represent more than 90\% of the genuine same-sign SM background to the SSDL+$2b$ final state. The fake background is made up by $t \bar t$ and $W b \bar b j$ events with one fake and one or more real leptons. OSDL events where a lepton charge mis-identification fakes a SSDL signal make up about $10\%$ of the background.  Our Fast Fake Simulation method could also be applied to the charge-flip background, but for simplicity we neglect this subdominant contribution and scale up the predicted fake lepton background by an additional 30\% to be conservative, which should amply compensate for our omission \cite{cmsSSDL2b, cmsSSDL2bLHC8}. 

All processes are calculated at lowest-order in \texttt{MadGraph 1.5.10}  \cite{madgraph}, showered in \texttt{Pythia 8.16} \cite{pythia6manual, pythia8}, and upscaled by the appropriate $K$-factors  \cite{Garzelli:2012bn, Campbell:2012dh, Czakon:2013goa, LHCHiggsCrossSectionWorkingGroup:2011ti, Dittmaier:2012vm, higgsNLO}. The events are analyzed in a \texttt{FastJet 3.0.2} \cite{fastjet} based \texttt{C++} code which aims to mimic closely the reconstruction of the CMS analysis, with anti-$k_T$(0.5) jet clustering \cite{antikt} and imposition of the same lepton-isolation conditions as \cite{cmsSSDL2b}. No jet-smearing is performed, but realistic efficiency curves for $b$-tagging and lepton-identification are folded in. The FakeSim was tuned to reproduce the expectations of \cite{cmsSSDL2b}. All the irreducible SM background expectations were successfully reproduced as well. 

The CMS SSDL+$2b$ analysis constructs signal regions in the $(\cancel{E}_T,H_T)$-plane, but this is not very useful for isolating the $t \bar t h$ signal, which was the reason for our dedicated collider study.  Since the signal is kinematically very similar to the dominant top-rich backgrounds, no one kinematic variable offers good discrimination. Due to the low rate of the $t \bar t h$ production process the measurement is likely statistics limited, and one may not want to perform harsh cuts beyond the preselection outlined above.

\begin{table}
\begin{center}
\begin{tabular}{m{1cm}c}
(a) &
\begin{tabular}{|l|c|c|c|}
\cline{2-4}
\multicolumn{1}{c|}{}&\multicolumn{3}{c|}{\# Events}\\
\hline
Process &7 TeV & 8 TeV & 8 Tev\\
& (5 $\ifb$)  & (5 $\ifb$) & (20 $\ifb$)\\
\hline \hline
$t\bar t W$  & 3.2    & 4.3 & 15.4\\
\hline
$t \bar t Z$  & 0.4  &  0.6 & 2.3  \\
\hline \hline
fakes (semileptonic $t \bar t$) & 4.3  &  9.9& 35.2\\
\hline
fakes (dileptonic $t \bar t j$)  & 0.97 & 2.3 & 9.4 \\ 
\hline
fakes ($W b \bar b j$) & 0.21 &  0.4 & 1.6   \\
\hline \hline
Total Background ($N_\mathrm{BG}$)& 9.18   & 17.5 & 64.0 \\
\hline \hline
$t \bar t h$ ($N_\mathrm{Sig}$)& 1.0  &  1.4 & 5.4 \\
\hline
\end{tabular}
\vspace{4mm}
\\
(b) & 
\begin{tabular}{|c| c | c | c | c | }
\cline{2-5}
\multicolumn{1}{c|}{}&\backslashbox{Signal}{BG} & 20\% & 30\%& 50\% \\
\hline
\multirow{2}{*}{LHC7}&10\% & 8.7 &  9.2 & 10.6  \\
&20\% & 9.3   & 9.9   &  11.7 \\
\hline\hline
\multirow{2}{*}{LHC8}&10\% &   5.0  & 6.7 & 9.7  \\
&20\% & 5.5   & 7.1   &  10.3 \\
\hline
\hline
LHC7 + LHC8&10\% &   5.7  & 6.3 &7.7  \\
($5\ifb + 5\ifb$)&20\% & 5.7   & 6.5   &  8.0 \\
\hline \hline
LHC7 + LHC8&10\% & 4.2 & 5.5 & 7.7 \\
($5\ifb + 20\ifb$)&20\% & 4.5 & 5.7 & 8.3 \\
\hline
\end{tabular}\vspace{4mm}
\\
(c) & 
\begin{tabular}{|c| c | c | c | c | }
\cline{2-5}
\multicolumn{1}{c|}{}&\backslashbox{Signal}{BG} & 20\% & 30\%& 50\% \\
\hline
\multirow{2}{*}{LHC7}&10\% &   10.2  &  10.7 & 11.7  \\
&20\% & 10.9   & 11.4   & 12.8 \\
\hline
\hline
\multirow{2}{*}{LHC8}&10\% &  6.1  &  7.6 & 10.7  \\
&20\% & 6.7   & 8.4   &  11.4 \\
\hline\hline
LHC7 + LHC8&10\%&   6.7  &  7.3 & 8.9  \\
($5\ifb + 5\ifb$)&20\% & 7.0   & 7.6   & 9.1 \\
\hline \hline
LHC7 + LHC8&10\% &5.3 &6.5 &8.9 \\
($5\ifb + 20\ifb$)&20\% & 5.6 &6.8 & 9.2\\
\hline
\end{tabular} 
\end{tabular} \vspace{-3mm}
\end{center}
\caption{
Result of $t\bar t h$-optimized cut and count analysis for LHC7 with 5$\ifb$, LHC8 with $20\ifb$ and a combination of LHC7 with 5$\ifb$ and LHC8 with $5\ifb$ and $20 \ifb$.  (a)~Events surviving the cuts in Table~\ref{t.cuts} without systematic uncertainty (statistical uncertainty of our predictions are negligible). (b)~Expected 95\% CL exclusion on signal strength $\mu_{t \bar t h}(b \bar b \ell^\pm \ell^\pm)$  assuming $N_\mathrm{obs} = N_\mathrm{BG}$ for different systematic uncertainties on the signal, as well as the irreducible SM background and fake lepton background (added in quadrature). (c)~Same as (b), but injecting the Standard Model signal: $N_\mathrm{obs} = N_\mathrm{BG} + N_\mathrm{SM}$. 
}
\label{t.LHC78colliderlimits}
\end{table}

To optimize our search strategy we identified six cuts that have high $ t \bar t h$ signal efficiency after preselection (except for (v), which has high background rejection):\vspace{-2mm}
\begin{enumerate}[(i)] \itemsep=-1mm
\item $\Delta R_\mathrm{SSDL} < \pi$, where $\Delta R_\mathrm{SSDL}$ is the separation between the two same-sign leptons;
\item $m_\mathrm{SSDL} < 175 \gev$;
\item $p_{T_{\ell 1}} < 250 \gev$, where the leptons are $p_T$-ordered;
\item $p_{T_{\ell 2}} < 100 \gev$;
\item The presence of a $3^\mathrm{rd}$ lepton; \vspace{-2mm} \vspace{2mm}
\item Veto any event where the leading light jet reconstructs a $W$-boson with one of the same-sign leptons, i.e. $|m_{j_1 \ell} - m_W| < 15 \gev$. 
\end{enumerate}
All possible AND/OR combinations of zero to three of these cuts were applied to our event samples resulting in 137 possible signal regions. For constraining $t \bar t h$ with $5\ifb$ of LHC7 data, the best cut was ``$(m_\mathrm{SSDL} < 175 \gev$) OR ($p_{T_{\ell 1}} < 250 \gev$ AND $\Delta R_\mathrm{SSDL} < \pi$)''. The chosen cuts for each analysis are summarized in \tref{cuts}. 
The result  is shown in \tref{LHC78colliderlimits}. 

Part (a) of this table shows the number of signal and background events surviving the optimally chosen set of cuts for each analysis. Statistical uncertainties of our Monte Carlo predictions are negligible, while different possible systematic uncertainties are illustrated in part (b) and (c), where the expected 95\% CL exclusions  on the signal strength  $\mu_{t \bar t h}(b \bar b \ell^\pm \ell^\pm)$  are collected. The search does not have SM sensitivity, but the expected signal strength limit is at least comparable to the already performed LHC7 $t \bar t h$ search in the $4b + \ell$ channel \cite{atlastthbb}, which achieved an observed (expected) signal strength limit of 13.1 (10.5).

We  repeated our LHC7 analysis for $\sqrt{s} = 8 \tev$ 20 $\ifb$, which is also summarized in  \tref{LHC78colliderlimits}. The optimal signal region is slightly modified: ``$(m_\mathrm{SSDL} < 175 \gev$) OR ($p_{T_{\ell 1}} < 250 \gev$ AND 3$^\mathrm{rd}$ lepton)''. With the systematic uncertainties on the irreducible SM and fake lepton backgrounds set at 50\% as in \cite{cmsSSDL2b, cmsSSDL2bLHC8}, the signal strength limit is 9.7, compared to the limit of 
\begin{eqnarray}
\mu_{t \bar t h}(2b2\gamma) \le  5.0 
\end{eqnarray}
 from the $t \bar t h$ search in the $2b 2\gamma$ channel \cite{cmstthgaga}. However, even a modest reduction in systematic uncertainties would make the signal strength limit  competitive. This provides clear motivation to the experimental collaborations to improve their background estimates. Furthermore, more sophisticated analysis techniques beyond the very simple cut and count we have employed could further improve discrimination.

The reach of our SSDL $t \bar t h$ search can be compared more directly to the $4b + \ell$ channel by obtaining exclusions for a combined LHC7 5$\ifb$ $+$ LHC8 $5\ifb$ dataset. The observed (expected) limit achieved by the corresponding CMS search \cite{cmstthbb} was 
\begin{eqnarray}
\mu_{t \bar t h}(4b + \ell)  \ \le \   5.8 \  (5.2).
\end{eqnarray}
For systematic uncertainties on the background (signal) in the range of 20 - 30\% (10 - 20\%) we can achieve an expected limit of 5.7 - 8.0, see \tref{LHC78colliderlimits}. The SSDL search channel is competitive, especially considering that our $y_\mathrm{top}$ constraints can be made independent of the poorly known bottom Yukawa by taking the ratio with a $Wh \rightarrow WWW^*$ measurement \cite{whtowww}, as explained in \sref{intro}. The relatively modest improvement that is achieved by combining with the full 20$\ifb$ LHC8 dataset underlines the necessity of reducing systematic backgrounds for the fake contribution.

 %%%%%%%%%%%%%%%%%%%%%%%%%%%%%%%%%%%%%%%%%%%%%%%%
\subsubsection{LHC with $\sqrt{s} = 14 \tev$}
 %%%%%%%%%%%%%%%%%%%%%%%%%%%%%%%%%%%%%%%%%%%%%%%%
\label{ss.lhc14}

\begin{table}
\begin{center}
\begin{tabular}{m{1cm}c}
(a) & 
\begin{tabular}{|l|l|}
\hline
Process & \# Events \\
\hline \hline
$t\bar t W$  & 4.3  \\
\hline
$t \bar t Z$ &  2.0  \\
\hline \hline
fakes (semileptonic $t \bar t$) & 0 \\
\hline
fakes (dileptonic $t \bar t j$)  & 13.7 (55.7) \\
\hline
fakes ($W b \bar b j$) & 0  \\
\hline \hline
Total Background ($N_\mathrm{BG}$)& 20.0 (62.0) \\
\hline \hline
$t \bar t h$ ($N_\mathrm{Sig}$) & 6.2   \\
\hline
\end{tabular}
\vspace{4mm}
\\
(b) & 
\begin{tabular}{| l | l | l | l | }\hline
\backslashbox{Signal}{BG} & 20\% & 30\%& 50\% \\
\hline
10\% & 2.1 (4.8) & 2.4 (6.4) & 3.1 (8.8)  \\
\hline
20\% & 2.2 (5.1)  & 2.5 (6.7) & 3.3  (9.5) \\
\hline
\end{tabular}
\vspace{4mm}
\\
(c) & 
\begin{tabular}{| l | l | l | l | }\hline
\backslashbox{Signal}{BG} & 20\% & 30\%& 50\% \\
\hline
10\% & 3.4 (5.8)  & 3.6 (7.4)  & 4.2 (10.0) \\
\hline
20\% & 3.7 (6.4)  & 3.9 (8.1) & 4.6 (10.8)  \\
\hline
\end{tabular} \vspace{-3mm}
\end{tabular}
\end{center}
\caption{
Result of $t\bar t h$-optimized cut and count analysis for LHC14 with 30$\ifb$ with optimistic (pessimistic) lepton fake rate calculation, see text for details. (a)~Events surviving the cuts in \tref{cuts} without systematic uncertainty (statistical uncertainty of our predictions are negligible). (b)~Expected 95\% CL exclusion on signal strength $\mu_{t \bar t h}(b \bar b \ell^\pm \ell^\pm)$  assuming $N_\mathrm{obs} = N_\mathrm{BG}$ for different systematic uncertainties on the signal, as well as the irreducible SM background and fake lepton background (added in quadrature). (c)~Same as (b), but for $N_\mathrm{obs} = N_\mathrm{BG} + N_\mathrm{SM}$. 
}
\label{t.LHC14colliderlimits}
\end{table}

As we have explained in \sref{fakesim}, the lack of calibration data for $\sqrt{s} = 14 \tev$ makes it difficult to derive definitive statements for the second run of the LHC. However, it is useful to demonstrate the $t \bar t h$ limit that can be achieved under fairly pessimistic assumptions. 

Carrying over the same analysis as for 7 and 8 TeV, the fake background now dominates the sample due to the rapidly growing top production cross sections. Much more aggressive cuts are therefore favored, in particular locking onto the tri-leptonic mode of $t \bar t h$ which eliminates most of the lepton fakes. The resulting signal strength limits are shown in \tref{LHC14colliderlimits} for the optimistic and pessimistic FakeSim scenario. Our cuts require a 3$^\mathrm{rd}$ lepton, $p_{T_{\ell1}} < 110 \gev$ and $m_\mathrm{SSDL} < 110\gev$, and likely exploit the fact that the lepton from Higgs decay often originates from a three-body decay and is hence softer, and more likely to be aligned with a leptonic top that the Higgs radiated off. 

Achieving Standard Model $t \bar t h$ coupling sensitivity is a plausible goal for the experimental collaborations with early LHC14 data, not only by reducing systematic uncertainties but also by utilizing more sophisticated multi-variate analysis techniques and/or splitting the signal into separate lepton flavor channels (both of which is beyond the scope of our minimally calibrated FakeSim). Measuring the $t \bar t h$ coupling may also motivate a dedicated lepton reconstruction that aims to minimize fakes, even at the cost of reducing signal efficiency by some $\mathcal{O}(1)$ factor.

 %%%%%%%%%%%%%%%%%%%%%%%%%%%%%%%%%%%%%%%%%%%%%%%%
 %%%%%%%%%%%%%%%%%%%%%%%%%%%%%%%%%%%%%%%%%%%%%%%%
\section{Higgs Couplings}
\label{s.Higgs} 
 %%%%%%%%%%%%%%%%%%%%%%%%%%%%%%%%%%%%%%%%%%%%%%%%
 %%%%%%%%%%%%%%%%%%%%%%%%%%%%%%%%%%%%%%%%%%%%%%%%

 %
%%%%%%%%%%%%%%%%%%%
\begin{figure}[tbp]
\begin{center}
\includegraphics[height=5.cm]{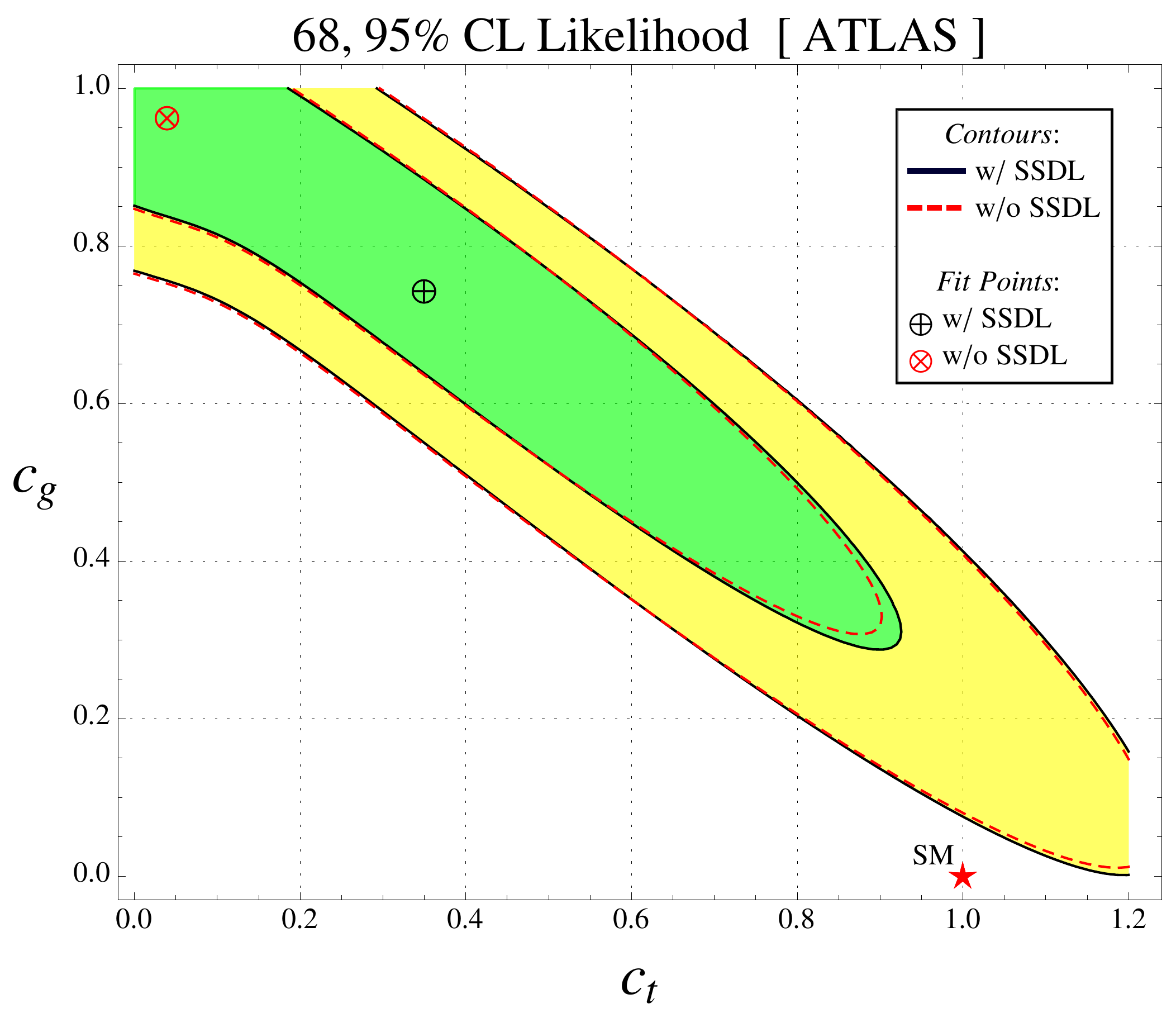}
\includegraphics[height=5.cm]{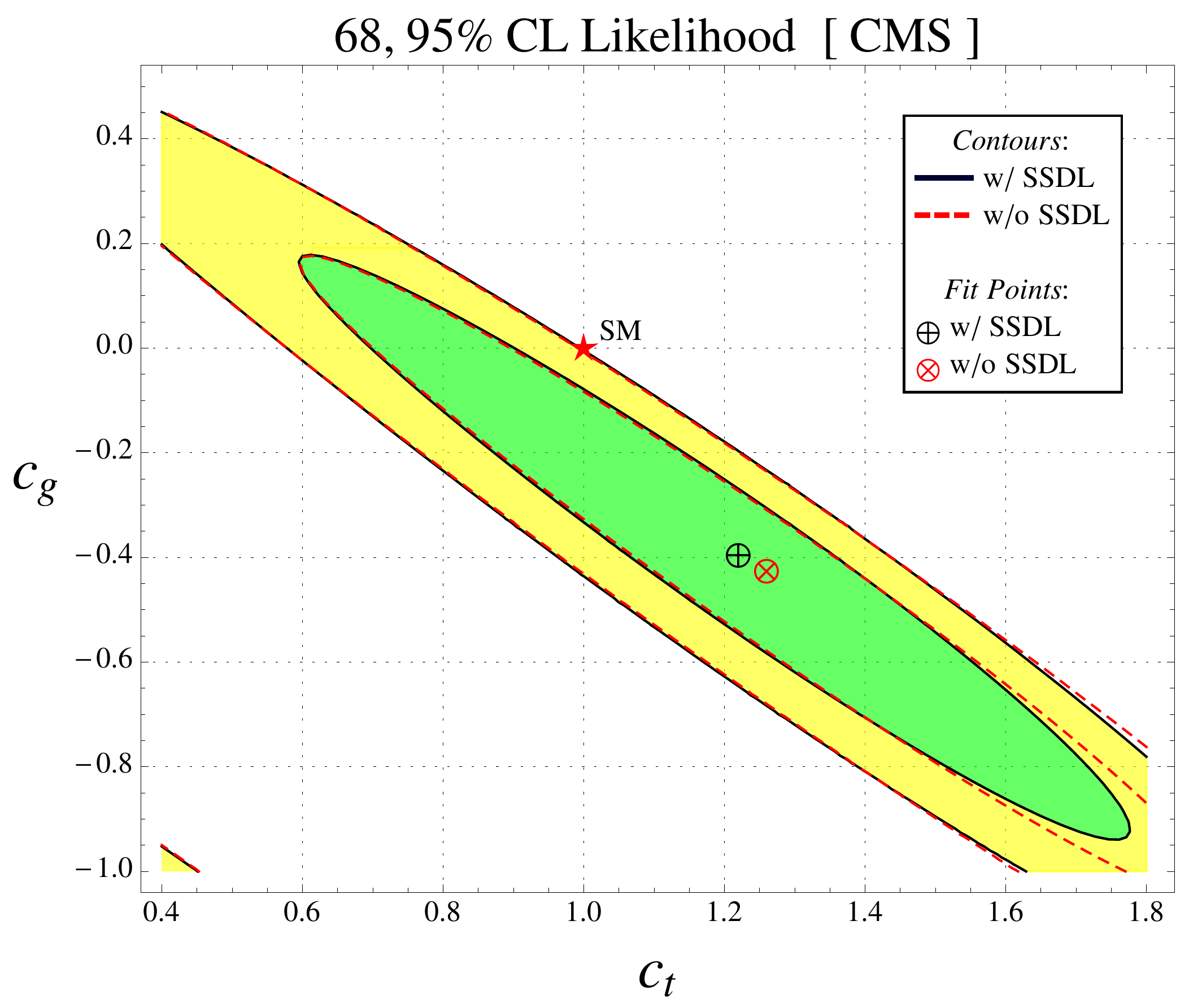}
\caption[]{\small
Higgs coupling to top quarks vs coupling to gluons via additional BSM contact operators; fits originate from 7 and 8 TeV runs and combine all channels as in \cite{AGrev}, with other Higgs couplings set to their SM values.  The relatively flat direction that exists in this plane and the preference for nonstandard couplings begins to find resolution with the assistance of $ttH$ with SSDL final states as calculated and described in  \tref{LHC78colliderlimits}.
\label{fig:gt}
}
\end{center}
\end{figure}
%%%%%%%%%%%%%%%%%%%
%

Accurate measurement of the top Yukawa is made difficult by the fact that this coupling enters either only through loops (as in the decay $h \to \gamma \gamma$ or production via gluon fusion) or through production cross-sections that constitute a minor fraction of Higgs events at the LHC.  Knowledge of this coupling  is crucial, however, as it 
begins to probe the part of the Higgs potential underlying the dominant contributor to the instability of the weak scale.  

The obvious utility of the channel studied here is that it grants {\it direct} access to the magnitude of this coupling, allowing us to address questions of the top quark's nature as well as to solidify our understanding of the global fit of Higgs couplings.  All the decays following $t \bar t h$ production involve only tree-level couplings, most of which are quite well determined.  The top decays proceed through gauge couplings, while the Higgs decay proceeds dominantly through its vector coupling, with a $\sim 25\%$ contribution from $h\rightarrow \tau \tau$. The vector coupling  at this point is known particularly well: accounting for LHC and LEP data, errors on this coupling are at the $\approx 5\%$ level (see for instance \cite{Falkowski} for a recent discussion). The tau Yukawa is currently known with approximately $40\%$ precision \cite{htautau}, but its decay is a subdominant component to our final state and its uncertainty will decrease  quickly with LHC14 data. We thus find it well within reason to fix the vector and tau couplings to their SM values in what follows. Additionally, we also set the total higgs width to its SM value. This is motivated by the $Wh\rightarrow WWW^*$ measurement \cite{whtowww}, which should establish sensitivity to SM signal strength with 30$\ifb$ of LHC14 data.  By taking the ratio of this channel's signal strength with our measurement we can effectively eliminate the dependence on total Higgs width (and thus the poorly known bottom Yukawa) without introducing dependence on any other poorly known couplings.

Regarding the issue of global fits, it is of particular interest to note that excesses in the Higgs decay to diphoton which appeared in early data and have persisted in the latest results of the ATLAS collaboration in fact lead to a very shallow direction of the global likelihood in a space spanned by the top coupling and new (BSM) contributions to gluon fusion.  Using an effective Lagrangian to capture these effects, we are interested in the following terms, working in unitary gauge:
\begin{eqnarray}
\Delta \mathcal L_{\rm eff}  =  \frac{h}{v} \times \left( c_t m_t \bar t t+ c_g G_{\mu \nu}^2\right),
\end{eqnarray}
i.e. $c_t$ denotes the top coupling to the Higgs normalized by its SM value as above, and $c_g$ parametrizes a new local operator contributing to gluon fusion with a coefficient that we normalize by the SM  loop contributions (e.g. $c_g=1$ corresponds to a doubling of the gluon fusion amplitude).
We illustrate the shallow direction of this space in Fig.~\ref{fig:gt}.

Clearly these two couplings have substantial overlap and require care to disentangle.  Appealing to the SSDL signature advocated here, along with complementing studies of $Wb \to t h$ \cite{Grojean, Mele} and $tth \to \gamma \gamma +X$ \cite{cmstthgaga}, is one way to begin a systematic discrimination of gluon fusion originating from SM versus BSM processes.

%%%%%%%%%%%%%%%%%%%
\begin{figure}[tbp]
\begin{center}
\includegraphics[height=5.cm]{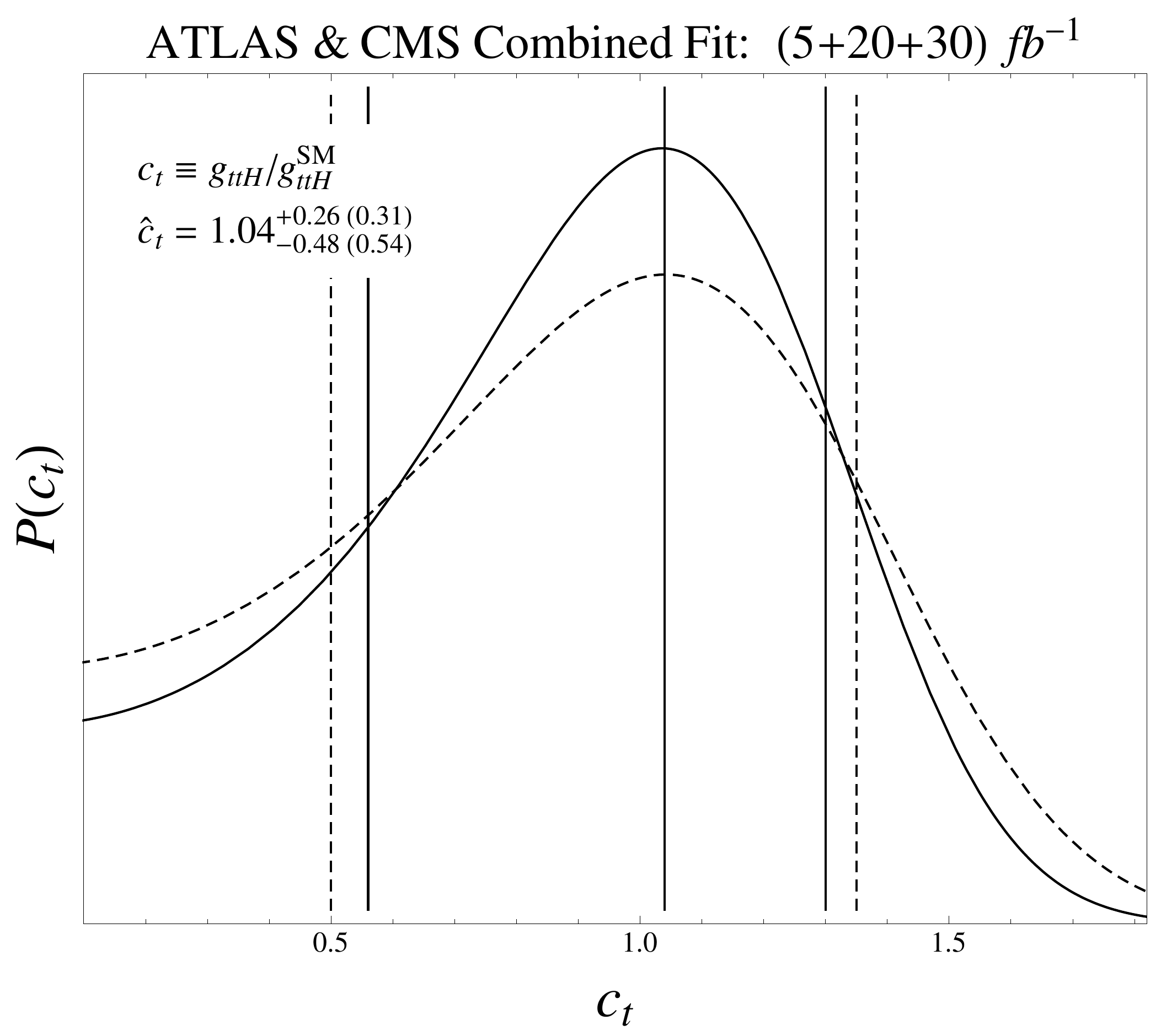}
\caption[]{\small
Fit of top coupling from the Higgs likelihood $P(c_t)$ coming exclusively from $ttH$ with SSDL in the final state after 55 $fb^{-1}$ from the 7, 8, and 14 TeV runs using the optimistic fake rate calculation for the latter.  Solid lines denote the likelihood and $\pm 1\sigma$ bands assuming systematic uncertainties of 20\% (10\%) on background (signal); dashed lines denote the same for systematic uncertainties of 30\% (20\%). 
\label{fig:ct}
}
\end{center}
\end{figure}
%%%%%%%%%%%%%%%%%%%
%

Ultimately we conclude that $tth$ production with SSDL final states can, with 30 $fb^{-1}$ from the 14 TeV LHC in addition to the data already on hand, determine the top coupling to within $\approx 40\%$ with background  uncertainties at the level of 30\% and  assuming optimal cuts as described in Section~\ref{s.collider}.
The likelihood here is constructed as a Poisson distribution in which theory predictions for signal and background events appear as variables $\theta_{\rm S,B}$ that are subsequently distributed with and marginalized over (truncated) Gaussians, $G$, according to their uncertainties $\sigma_{\rm S,B}$; i.e.
\begin{eqnarray}
P(c_t)  =  N \int d \theta_{\rm S} \, d \theta_{\rm B} \,
 (\theta_{\rm B} + c_t^2 \, \theta_{\rm S})^{n_{\rm obs}} 
 \, e^{-(\theta_{\rm B} + c_t^2 \, \theta_{\rm S})} \\ \nonumber
 \times
 \, G (\theta_{\rm S},\sigma_{\rm S}) \, G(\theta_{\rm B},\sigma_{\rm B}). \hspace{-0.5cm}
\end{eqnarray}
Here $N$ represents a normalization factor specific to the parameter space of interest, and the quantity $n_{\rm obs}$ corresponds simply to $n_{\rm B} + c_t^2 n_{\rm S}$, with $n_{\rm S,B}$ the central values of the signal and background distributions.  The resulting likelihoods are shown in ~\ref{fig:ct}.

\vspace{4mm}

 %%%%%%%%%%%%%%%%%%%%%%%%%%%%%%%%%%%%%%%%%%%%%%%%
 %%%%%%%%%%%%%%%%%%%%%%%%%%%%%%%%%%%%%%%%%%%%%%%%
\section{Conclusions}
\label{s.conclusions}
 %%%%%%%%%%%%%%%%%%%%%%%%%%%%%%%%%%%%%%%%%%%%%%%%
 %%%%%%%%%%%%%%%%%%%%%%%%%%%%%%%%%%%%%%%%%%%%%%%%

We have demonstrated that the SSDL $+2b$ channel presents an attractive experimental opportunity to measure the theoretically important top Yukawa coupling. The main experimental challenge lies in reducing both the size and systematic uncertainty of the fake lepton contamination, which is a plausible goal to achieve for the early run of LHC14. Standard-model like sensitivity to  $t \bar t h$ production is possible with $30 \ifb$ of data. This measurement, apart from improving overall Higgs coupling fits by providing an independent additional channel, has the additional advantage that its sensitivity to the top Yukawa coupling is not convoluted with any loop couplings or other poorly determined Yukawas after removing the total higgs width dependence using \cite{whtowww}. A precise SSDL$+2b$ measurement also has the potential to resolve any remaining ambiguity between the top-loop and any additional BSM contribution to the $hgg$ coupling. 

We developed a new method of calculating lepton fake rates which is computationally efficient, grounded in well-motivated approximations and tunable to data. This allowed us to perform our early feasibility study of $t \bar t h \rightarrow \ell^\pm \ell^\pm + 2b$, but the lack of available calibration data likely means our LHC14 sensitivity projections might be substantially reduced by altering lepton identification criteria.  The Fast FakeSim will be developed further in \cite{fakesim} and can be applied to many other particles which have `fake' backgrounds such as tau leptons and photons.

\subsection*{Acknowledgements}
We thank Agnes Taffard and Dmitri Tsybychev  for helpful discussions, and are grateful to Jared Evans for drawing our attention to the potential significance of contamination in our signal from SSDL originating in $h \to \tau \tau$.  Finally we thank Tim Cohen and Mariangela Lisanti for reading the draft and providing useful feedback.
D.C. is supported in part by the National Science Foundation under Grant PHY-0969739. 
J.G. is supported by the ERC Advanced Grant No.~267985. 
{\em Electroweak Symmetry Breaking, Flavour and  Dark Matter: One Solution for Three Mysteries (DaMeSyFla).}
J.G.W. is supported by the US Department of Energy under contract number DE-AC02-76SF00515.


\begin{thebibliography}{99}




\bibitem{ATLASdisc}
  G.~Aad {\it et al.}  [ATLAS Collaboration],
  %``Observation of a new particle in the search for the Standard Model Higgs boson with the ATLAS detector at the LHC,''
  Phys.\ Lett.\ B {\bf 716}, 1 (2012)
  [arXiv:1207.7214 [hep-ex]].
  %%CITATION = ARXIV:1207.7214;%%





\bibitem{CMSdisc}
  S.~Chatrchyan {\it et al.}  [CMS Collaboration],
  %``Observation of a new boson at a mass of 125 GeV with the CMS experiment at the LHC,''
  Phys.\ Lett.\ B {\bf 716}, 30 (2012)
  [arXiv:1207.7235 [hep-ex]].
  %%CITATION = ARXIV:1207.7235;%%





\bibitem{ATLASprop}
  [ATLAS Collaboration],
  %``Combined coupling measurements of the Higgs-like boson with the ATLAS detector using up to 25 fb$^{-1}$ of proton-proton collision data,''
  ATLAS-CONF-2013-034.
  %%CITATION = ATLAS-CONF-2013-034;%%





\bibitem{CMSprop}
  [CMS Collaboration],
  %``Combination of standard model Higgs boson searches and measurements of the properties of the new boson with a mass near 125 GeV,''
  CMS-PAS-HIG-12-045.
  %%CITATION = CMS-PAS-HIG-12-045;%%


\bibitem{cmstthgaga}
[CMS Collaboration]
%``Search for top-antitop-Higgs production in events where $H \rightarrow \gamma \gamma$ at $\sqrt{s} = 8 \tev$  collisions'', 
CMS-HIG-13-015.

\bibitem{cmstthbb}
  S.~Chatrchyan {\it et al.}  [CMS Collaboration],
  %``Search for the standard model Higgs boson produced in association with a top-quark pair in pp collisions at the LHC,''
  JHEP {\bf 1305}, 145 (2013)
  [arXiv:1303.0763 [hep-ex]].
  %%CITATION = ARXIV:1303.0763;%%
  %4 citations counted in INSPIRE as of 13 Jun 2013
  
\bibitem{atlastthbb} 
  [ATLAS Collaboration],
  %``Search for the Standard Model Higgs boson produced in association with top quarks in proton-proton collisions at Ãs = 7 TeV using the ATLAS detector,''
  ATLAS-CONF-2012-135.
  %%CITATION = ATLAS-CONF-2012-135;%%



  \bibitem{whtowww}
  CMS Collaboration, Hig13009TWiki.
  
  
  
\bibitem{oldtthww} 
  F.~Maltoni, D.~L.~Rainwater and S.~Willenbrock,
  %``Measuring the top quark Yukawa coupling at hadron colliders via $t\bar{t}H,H\to W^+W^-$,''
  Phys.\ Rev.\ D {\bf 66}, 034022 (2002)
  [hep-ph/0202205].
  %%CITATION = HEP-PH/0202205;%%
  %52 citations counted in INSPIRE as of 28 Jun 2013
  
  

 \bibitem{cmsSSDL2b} 
  S.~Chatrchyan {\it et al.}  [CMS Collaboration],
  %``Search for new physics in events with same-sign dileptons and $b$-tagged jets in $pp$ collisions at $\sqrt{s}=7$ TeV,''
  JHEP {\bf 1208}, 110 (2012)
  [arXiv:1205.3933 [hep-ex]].
  %%CITATION = ARXIV:1205.3933;%%
  %55 citations counted in INSPIRE as of 13 Jun 2013
  
  
  
  \bibitem{cmsSSDL2bLHC8} 
  S.~Chatrchyan {\it et al.}  [CMS Collaboration],
  %``Search for new physics in events with same-sign dileptons and $b$ jets in $pp$ collisions at $\sqrt{s}=8$ TeV,''
  JHEP {\bf 1303}, 037 (2013)
  [arXiv:1212.6194 [hep-ex]].
  %%CITATION = ARXIV:1212.6194;%%
  %10 citations counted in INSPIRE as of 18 Jun 2013
  
  
  \bibitem{atlasSSDL}
  [ATLAS Collaboration],
  %``Search for strongly produced superpartners in final states with two same sign leptons with the ATLAS detector using 21 fb-1 of proton-proton collisions at sqrt(s)=8 TeV.,''
  ATLAS-CONF-2013-007;
  %%CITATION = ATLAS-CONF-2013-007;%%
  %4 citations counted in INSPIRE as of 19 Jun 2013
  [ATLAS Collaboration],
  %``Measurement of the production cross section of prompt $J/\psi$ mesons in association with a $W^{\pm}$ boson in $pp$ collisions at $\sqrt{s}=7$ TeV,''
  ATLAS-CONF-2013-042.
  %%CITATION = ATLAS-CONF-2013-042;%%
  
  
  
  \bibitem{htautau}
  CMS Collaboration, CMS PAS HIG-13-004
  

  \bibitem{Falkowski}
  A.~Falkowski, F.~Riva and A.~Urbano,
  %``Higgs At Last,''
  arXiv:1303.1812 [hep-ph].
  %%CITATION = ARXIV:1303.1812;%%
  
  
  
\bibitem{fakesim}
D.~Curtin,  J.~Wacker,
``Fast Calculation of Fake Backgrounds'',
to appear shortly. 



\bibitem{cmsSSDL2j} 
  S.~Chatrchyan {\it et al.}  [CMS Collaboration],
  %``Search for new physics with same-sign isolated dilepton events with jets and missing transverse energy,''
  Phys.\ Rev.\ Lett.\  {\bf 109}, 071803 (2012)
  [arXiv:1205.6615 [hep-ex]].
  %%CITATION = ARXIV:1205.6615;%%
  %35 citations counted in INSPIRE as of 13 Jun 2013
  
  
\bibitem{effcurvelinear}
Jeffrey Berryhill, ``Electron Detection at CMS'', Seminar given on 3. August 2009

\bibitem{cdftransferfnmeasurement}
Martin Griffiths, Giulia Manca, Beate Heinemann, ``Studying the fake lepton rate for the trilepton analysis'', Seminar given on behalf of the CDF collaboration. 
  
  
\bibitem{madgraph} 
  J.~Alwall, M.~Herquet, F.~Maltoni, O.~Mattelaer and T.~Stelzer,
  %``MadGraph 5 : Going Beyond,''
  JHEP {\bf 1106}, 128 (2011)
  [arXiv:1106.0522 [hep-ph]].
  %%CITATION = ARXIV:1106.0522;%%
  %602 citations counted in INSPIRE as of 23 May 2013
  
  
  
\bibitem{pythia6manual} 
  T.~Sjostrand, S.~Mrenna and P.~Z.~Skands,
  %``PYTHIA 6.4 Physics and Manual,''
  JHEP {\bf 0605}, 026 (2006)
  [hep-ph/0603175].
  %%CITATION = HEP-PH/0603175;%%
  
  
  
  \bibitem{pythia8}
  T.~Sjostrand, S.~Mrenna and P.~Z.~Skands,
  %``A Brief Introduction to PYTHIA 8.1,''
  Comput.\ Phys.\ Commun.\  {\bf 178}, 852 (2008)
  [arXiv:0710.3820 [hep-ph]].
  %%CITATION = ARXIV:0710.3820;%%
  
    
  

\bibitem{Garzelli:2012bn} 
  M.~V.~Garzelli, A.~Kardos, C.~G.~Papadopoulos and Z.~Trocsanyi,
  %``t $\bar{t}$ $W^{+-}$ and t $\bar{t}$ Z Hadroproduction at NLO accuracy in QCD with Parton Shower and Hadronization effects,''
  JHEP {\bf 1211}, 056 (2012)
  [arXiv:1208.2665 [hep-ph]].
  %%CITATION = ARXIV:1208.2665;%%
  %13 citations counted in INSPIRE as of 17 Jun 2013
  
\bibitem{Campbell:2012dh} 
  J.~M.~Campbell and R.~K.~Ellis,
  %``$t \bar{t} W^{+-}$ production and decay at NLO,''
  JHEP {\bf 1207}, 052 (2012)
  [arXiv:1204.5678 [hep-ph]].
  %%CITATION = ARXIV:1204.5678;%%
  %48 citations counted in INSPIRE as of 17 Jun 2013
  
  \bibitem{Czakon:2013goa} 
  M.~Czakon, P.~Fiedler and A.~Mitov,
  %``The total top quark pair production cross-section at hadron colliders through O(alpha_S^4),''
  arXiv:1303.6254 [hep-ph].
  %%CITATION = ARXIV:1303.6254;%%
  %14 citations counted in INSPIRE as of 17 Jun 2013
  
  
  \bibitem{LHCHiggsCrossSectionWorkingGroup:2011ti}
  LHC Higgs Cross Section Working Group, S.~Dittmaier, C.~Mariotti, G.~Passarino, and R.~Tanaka (Eds.), 
  {\sl Handbook of LHC Higgs Cross Sections: 1. Inclusive Observables}, 
  CERN-2011-002 (CERN, Geneva, 2011), {\tt arXiv:1101.0593 [hep-ph]}.

\bibitem{Dittmaier:2012vm}
  LHC Higgs Cross Section Working Group, S.~Dittmaier, C.~Mariotti, G.~Passarino, and R.~Tanaka (Eds.), 
  {\sl Handbook of LHC Higgs Cross Sections: 2. Differential Distributions}, 
  CERN-2012-002 (CERN, Geneva, 2012), {\tt arXiv:1201.3084 [hep-ph]}.


\bibitem{higgsNLO}
  W.~Beenakker, S.~Dittmaier, M.~Kramer, B.~Plumper, M.~Spira and P.~M.~Zerwas,
  %``Higgs radiation off top quarks at the Tevatron and the LHC,''
  Phys.\ Rev.\ Lett.\  {\bf 87}, 201805 (2001)
  [hep-ph/0107081];
  %%CITATION = HEP-PH/0107081;%%
  %251 citations counted in INSPIRE as of 28 Jun 2013
  W.~Beenakker, S.~Dittmaier, M.~Kramer, B.~Plumper, M.~Spira and P.~M.~Zerwas,
  %``NLO QCD corrections to t anti-t H production in hadron collisions,''
  Nucl.\ Phys.\ B {\bf 653}, 151 (2003)
  [hep-ph/0211352];
  %%CITATION = HEP-PH/0211352;%%
  %212 citations counted in INSPIRE as of 28 Jun 2013
  L.~Reina and S.~Dawson,
  %``Next-to-leading order results for t anti-t h production at the Tevatron,''
  Phys.\ Rev.\ Lett.\  {\bf 87}, 201804 (2001)
  [hep-ph/0107101];
  %%CITATION = HEP-PH/0107101;%%
  %130 citations counted in INSPIRE as of 28 Jun 2013
  S.~Dawson, C.~Jackson, L.~H.~Orr, L.~Reina and D.~Wackeroth,
  %``Associated Higgs production with top quarks at the large hadron collider: NLO QCD corrections,''
  Phys.\ Rev.\ D {\bf 68}, 034022 (2003)
  [hep-ph/0305087].
  %%CITATION = HEP-PH/0305087;%%
  %119 citations counted in INSPIRE as of 28 Jun 2013





\bibitem{fastjet}
    M.~Cacciari and G.~P.~Salam,
  %``Dispelling the $N^{3}$ myth for the $k_t$ jet-finder,''
  Phys.\ Lett.\ B {\bf 641}, 57 (2006)
  [hep-ph/0512210];
  %%CITATION = HEP-PH/0512210;%%
    M.~Cacciari, G.~P.~Salam and G.~Soyez,
  %``FastJet user manual,''
  Eur.\ Phys.\ J.\ C {\bf 72}, 1896 (2012)
  [arXiv:1111.6097 [hep-ph]].
  %%CITATION = ARXIV:1111.6097;%%

\bibitem{antikt} 
  M.~Cacciari, G.~P.~Salam and G.~Soyez,
  %``The Anti-k(t) jet clustering algorithm,''
  JHEP {\bf 0804}, 063 (2008)
  [arXiv:0802.1189 [hep-ph]].
  %%CITATION = ARXIV:0802.1189;%%

\bibitem{AGrev}
  A.~Azatov and J.~Galloway,
  %``Electroweak Symmetry Breaking and the Higgs Boson: Confronting Theories at Colliders,''
  Int.\ J.\ Mod.\ Phys.\ A {\bf 28}, 1330004 (2013)
  [arXiv:1212.1380 [hep-ph]].
  %%CITATION = ARXIV:1212.1380;%%


\bibitem{Grojean} 
  M.~Farina, C.~Grojean, F.~Maltoni, E.~Salvioni and A.~Thamm,
  %``Lifting degeneracies in Higgs couplings using single top production in association with a Higgs boson,''
  JHEP {\bf 1305}, 022 (2013)
  [arXiv:1211.3736 [hep-ph]].
  %%CITATION = ARXIV:1211.3736;%%


  
\bibitem{Mele}
  S.~Biswas, E.~Gabrielli and B.~Mele,
  %``Single top and Higgs associated production as a probe of the Htt coupling sign at the LHC,''
  JHEP {\bf 1301}, 088 (2013)
  [arXiv:1211.0499 [hep-ph]].
  %%CITATION = ARXIV:1211.0499;%%
  
  
\end{thebibliography}
\end{document}